\documentclass[10pt,letterpaper,twocolumn]{article}
\usepackage{ol2}
\usepackage[draft]{hyperref}
\usepackage{amsmath}

\begin{document}


\twocolumn[
\title{Partially coherent sources  which produce the same far zone optical force as a laser beam}

\author{Juan Miguel  Au\~{n}\'{o}n$^{*}$ and Manuel  Nieto-Vesperinas  }

\address{Instituto de Ciencia de Materiales de Madrid, Consejo Superior
de Investigaciones Cientificas, Campus de Cantoblanco, Madrid 28049,
Spain.
\\ 
$^*$jmaunon@icmm.csic.es}

\begin{abstract}
On applying a theorem previously derived by Wolf and Collett, we demonstrate that  partially coherent  Gaussian Schell model fluctuating sources (GSMS) produce exactly the same optical forces as a fully coherent laser beam.  {\it We also show that this kind of sources helps to control the light-matter interaction in biological samples which are very sensitive to thermal heating induced by higher power intensities; and hence the invasiveness of the manipulation}. This is a consequence of the fact that the same photonic force can be obtained with a low intensity GSMS as with a high intensity laser beam.
\end{abstract}

\ocis{030.1640,140.7010,260.0260}
]

\noindent 
\maketitle

\maketitle

The nanomanipulation of submicrometer particles or biological samples
is a multidisciplinary area of great interest in science. Since the seminal works by Ashkin \cite{Ashkin1970}, many studies have dealt with optical traps  either within the Rayleigh limit of resolution \cite{Ashkin1987science, grier2003revolution,ChaumetOL}, or beyond \cite{Chaumet2002PRL, Kall2002PRL, Arias2002modeling, Arias2003attractive,  Valdivia2012forces}, nevertheless, and apart from a few recent works \cite{wang2007effects, korotkova2009forces, aunon2012photonic, aunon2012opticalforces},  most attention has  been paid  to coherent light.   The now widely used optical tweezers generally employ $\text{TEM}_{00}$ laser  Gaussian beams; and still there are very few studies on the influence of their  statistical properties as regards their use as optical manipulators \cite{korotkova2009forces}. We next consider  the wide variety of  Gaussian Schell model sources (GSMS) \cite{mandel1995optical, Roychowdhury2005Realizability}. 

In this regard, an equivalence theorem (ET)  was demonstrated by Wolf and Collett more than twenty years ago \cite{Wolf1978theorem}  according to which  a fully coherent laser is not necessary to produce a highly directional intensity distribution, thus proving that some partially coherent sources, known as those above mentioned GSMS, fulfil this property.

The aim of this letter is to demonstrate that, as consequence of the ET, {\it a partially coherent fluctuating source produces the same optical force as laser beam}. In addition, this may be done even with a much lower peak intensity. We think that this fact opens  a new door in the area of optical manipulation because, as we will show, it is not necessary to have a  rather intense spatially coherent beam in order to create a potential well with an equilibrium position for a particle in the optical trap.

It is well-known that the mechanical action from a light field at frequency $\omega$ with electric vector $\bf{E(\textbf{r},\omega)}$ on a dipolar particle in vacuum with permittivity $\varepsilon_0$, i.e., that whose scattering cross section can be fully expressed in terms of the two first Mie coefficients $a_1$ and $b_1$,  can be written as a sum of a gradient (conservative) and a scattering plus curl (non-conservative) force \cite{nieto2004near,wong2006gradient,albaladejo2009Scattering, nieto2010optical}, thus 
\begin{eqnarray}
F_{i}\left(\mathbf{r},\omega\right) & = & F_{i}^{cons}\left(\mathbf{r},\omega\right)+F_{i}^{nc}\left(\mathbf{r},\omega\right)\nonumber \\ 
 & = & \frac{\varepsilon_0}{4}\text{Re}\alpha_e\partial_{i}\left\langle E_{j}^{*}\left(\mathbf{r},\omega\right)E_{j}\left(\mathbf{r},\omega\right)\right\rangle \nonumber\\
 &  & +\frac{\varepsilon_0}{2}\text{Im}\alpha_e\left\langle E_{j}^{*}\left(\mathbf{r},\omega\right)\partial_{i}E_{j}\left(\mathbf{r},\omega\right)\right\rangle , \label{force}
\end{eqnarray}
$(i,j)=(x,y,z)$ and $\alpha_e$ is the electric polarizabity of the particle. The electric field
can be written in terms of an angular spectrum  of plane waves $\textbf{e}\left(k\mathbf{s}_{\perp},\omega\right)$ \cite{conjug,mandel1995optical,nietolibro}
\begin{equation}
\mathbf{E}\left(\mathbf{r},\omega\right)=\int\mathbf{e}\left(k\mathbf{s}_{\perp},\omega\right)e^{ik\mathbf{s}\cdot\mathbf{r}}d^{2}\mathbf{s}_{\perp},\label{angesp}
\end{equation}
where $k=\omega/c$ and $c$ the speed of light in vacuum  The unit vector $\mathbf{s}=\left(\mathbf{s}_{\perp},s_{z}\right)$, is such that $\textbf{s}_{\perp}=(s_x,s_y)$, $s_{z}=\sqrt{1-s_{\perp}^{2}}$, ($|\textbf{s}|^{2}_{\perp} \leq 1$) and $s_{z}=i\sqrt{s_{\perp}^{2}-1}$, ($|\textbf{s}|^{2}_{\perp}>1$) for propagating and evanescent waves, respectively. Thus, using Eq.(\ref{angesp})  the force components can be written in terms of the statistical properties of the source as \cite{aunon2012photonic}

\begin{eqnarray}
F_{i}^{cons}\left(\mathbf{r},\omega\right) & =-i\frac{k}{4}\text{Re}\alpha_{e}\iint_{-\infty}^{\infty}\textrm{Tr}\mathcal{A}_{jk}^{(e)}\left(k\mathbf{s}_{\perp},k\mathbf{s'}_{\perp}\omega\right)\nonumber \\
 & \times\left(s_{i}^{*}-s'_{i}\right)e^{-ik\left(\mathbf{s}^{*}-\mathbf{s'}\right)\cdot\mathbf{r}}d^{2}{\bf s}_{\perp}d^{2}{\bf s}'_{\perp},
\label{forcecons} 
\end{eqnarray}
 
\begin{eqnarray}
F_{i}^{nc}\left(\mathbf{r},\omega\right) & =\frac{1}{2}\text{Im}\alpha_{e}\text{Im}\left\{ ik\iint_{-\infty}^{\infty}\textrm{Tr}\mathcal{A}_{jk}^{(e)}\left(k\mathbf{s}_{\perp},k\mathbf{s'}_{\perp}\omega\right)\right.\nonumber \\
 & \times\left.s'_{i}e^{-ik\left(\mathbf{s}^{*}-\mathbf{s'}\right)\cdot\mathbf{r}}d^{2}{\bf s}_{\perp}d^{2}{\bf s}'_{\perp}\right\},
 \label{forcenc} 
\end{eqnarray}
where $\textrm{Tr}$ denotes the trace of the electric
angular correlation tensor $\mathcal{A}_{jk}^{(e)}\left(k\mathbf{s}_{\perp},k\mathbf{s'}_{\perp},\omega\right)=\left\langle e_{j}^{*}(k\mathbf{s}_{\perp},\omega)e_{k}(k\mathbf{s'}_{\perp},\omega)\right\rangle $.
Notice, that Eqs. (\ref{forcecons})-(\ref{forcenc})  relate the optical force with the statistical properties  of the source trough this correlation tensor and is valid for any  GSMS, or in particular  for a homogeneous or quasi-homogeneous source \cite{mandel1995optical}. 

Addressing then a planar GSMS,   its cross spectral
density tensor $W_{ij}^{(0)}\left(\mathbf{\boldsymbol{\rho}}_{1},\mathbf{\boldsymbol{\rho}}_{2},\omega\right)=\left\langle E_{i}^{*}\left(\boldsymbol{\rho}_{1}\right)E_{j}\left(\boldsymbol{\rho}_{2}\right)\right\rangle $ at the plane  $z=0$ of the source, is given by \cite{mandel1995optical}
\begin{eqnarray}
& &W_{ij}^{(0)}\left(\mathbf{\boldsymbol{\rho}}_{1},\mathbf{\boldsymbol{\rho}}_{2},\omega\right) \nonumber \\
&=&\sqrt{S_{i}^{(0)}\left(\mathbf{\boldsymbol{\rho}}_{1},\omega\right)}\sqrt{S_{j}^{(0)}\left(\mathbf{\boldsymbol{\rho}}_{2},\omega\right)} 
\mu_{ij}^{(0)}\left(\mathbf{\boldsymbol{\rho}}_{2}-\mathbf{\boldsymbol{\rho}}_{1},\omega\right)
\end{eqnarray}
where $S^{(0)}$ and  $\mu^{(0)}$ are the spectral density and  the spectral degree of coherence of the source, respectively. For this type of sources these quantities are both Gaussian, i.e.,
\begin{equation}
S_{i}\left(\mathbf{\boldsymbol{\rho}},\omega\right)=A_{i}\text{exp}{[-\rho^{2}/(2\sigma_{s,i}^{2}(\omega))]} ,
\end{equation}
\begin{equation}
\mu_{ij}\left(\mathbf{\boldsymbol{\rho}}_{2}-\mathbf{\boldsymbol{\rho}}_{1},\omega\right) = B_{ij}\text{exp}[-|\boldsymbol{\rho}_{2}-\boldsymbol{\rho}_{1}|^{2}/(2\sigma_{g,ij}^{2}(\omega))].
\end{equation}
The parameter $A_i$ is the maximum of the spectral intensity at $\boldsymbol{\rho}=0$, whereas $B_{ij}$ is $|B_{ij}|=1$ if $i=j$ and  $|B_{ij}| \leq1$ if $i\neq j$.

The widths $\sigma(\omega)_{s,i}$ and $\sigma(\omega)_{g,ij}$ are usually known as the spot size and the correlation or spatial coherence length of the source,  respectively. Notice that these parameters cannot be chosen arbitrarily, they have to fulfil a series of conditions \cite{Roychowdhury2005Realizability}. In this work, for simplicity, we restrict ourselves to the case in which the field is completely polarized, i.e., the  degree of polarization is equal to 1 \cite{wolf2007introduction}, or equivalently, the electric field only fluctuates in one direction (for example in the $x-$direction). At fixed frequency and in order to simplify the notation, in what follows we shall write $\sigma_{i,s}$ and $\sigma_{ij,g}$ without  $\omega$ dependence  nor  the Cartesian subindex, understanding that we are dealing with the $x-$component of the electric field .  

Now we turn to  analytically calculate the optical forces in the far-zone. To this end we need the angular correlation tensor at the plane of the source. This  has been previously calculated, thus the trace of the angular correlation tensor  reads as \cite{mandel1995optical}  
\begin{eqnarray}
&&\text{Tr}{\cal A}_{ij}(k\mathbf{s}_{\perp,1},k\mathbf{s}_{\perp,2})\simeq{\cal A}_{xx}(k\mathbf{s}_{\perp,1},k\mathbf{s}_{\perp,2})\nonumber\\
&=&k^{4}\frac{A}{(4\pi)^{2}(a^{2}-b^{2})}e^{-(\alpha k^{2}\mathbf{s}_{\perp,1}^{2}+\alpha k^{2}\mathbf{s}_{\perp,2}^{2}-2k^{2}\beta\mathbf{s}_{\perp,1}\mathbf{s}_{\perp,2})},
\label{trace_angular} 
\end{eqnarray}
where $a=1/\left(4\sigma_{s}^{2}\right)+1/\left(2\sigma_{g}^{2}\right)$, $b=1/\left(2\sigma_{g}^{2}\right)$, $\alpha=a/4(a^{2}-b^{2})$  and $\beta=b/4(a^{2}-b^{2})$

One of the most important  characteristics of these GSMS sources
is that the behavior of the emitted  field can be beam-like. To ensure  this in  the far-zone, the following necessary and sufficient conditions have to be fulfilled \cite{mandel1995optical}
\begin{equation}
\frac{1}{(2\sigma_{s})^{2}}+\frac{1}{\sigma_{g}^{2}}\ll\frac{2\pi^{2}}{\lambda^{2}}.
\label{beam_condition}
\end{equation}

Next, in order to obtain the force in SI units, we redefine the parameter $A$ as $A/(\varepsilon_0c)$, where $A$ is the peak intensity of the source in $W/m^2$.
Substituting Eq. (\ref{trace_angular}) into Eqs. (\ref{forcecons})-(\ref{forcenc}), approximating $s_z \simeq 1-1/2s_\perp^2$, and after a long tedious but straightforward calculation, one derives the different components of the force. Then, performing the ${\bf s}_{\perp}$ and $ {\bf s}'_{\perp}$ integrations, the conservative components finally are 
\begin{equation}
F_{x,y}^{cons}=-\text{Re}\alpha_e\frac{A}{\varepsilon_0 c}\frac{1}{4\sigma_{s}^{2}\Delta(z)^{4}}e^{-\frac{\boldsymbol{\rho}^{2}}{2(\sigma_{s}\Delta(z))^{2}}}(x,y)
\label{Fx_cons}
\end{equation}
and 
\begin{equation}
F_{z}^{cons}=\text{Re}\alpha_{e}\frac{Az}{4k^{2}\sigma_{s}^{4}\delta^{2}\Delta(z)^{6}\varepsilon_0c}\left(\rho^{2}-2\sigma_{s}^{2}\Delta(z)^{2}\right)e^{-\frac{\boldsymbol{\rho}^{2}}{2(\sigma_{s}\Delta(z))^{2}}}.
\label{Fz_cons}
\end{equation}
On the other hand the non-conservative forces read
\begin{equation}
F_{x,y}^{nc}=\text{Im}\alpha_e\frac{A}{\varepsilon_0 c}\frac{z}{2k\sigma_{s}^{2}\delta^{2}\Delta(z)^{4}}e^{-\frac{\boldsymbol{\rho}^{2}}{2(\sigma_{s}\Delta(z))^{2}}}(x,y)
\label{Fx_nc}
\end{equation}
and
\begin{eqnarray}
&&F_{z}^{nc} \nonumber\\
&=&\text{Im}\alpha_{e}\frac{A}{2k^{7}\sigma_{s}^{4}\delta^{4}\Delta(z)^{6}\varepsilon_0c}\left[\frac{1}{2}k^{8}\sigma_{s}^{4}\delta^{4}\Delta(z)^{4}-\alpha k^{6}\sigma_{s}^{2}\delta^{2}\Delta(z)^{2}\right.\nonumber\\
&&\left.+\left(\frac{k^{4}}{2^{4}}\delta^{4}-\frac{k^{2}}{4}z^{2}\right)k^{2}\rho^{2}\right]e^{-\frac{\boldsymbol{\rho}^{2}}{2(\sigma_{s}\Delta(z))^{2}}},
\label{Fz_nc}
\end{eqnarray}
where $1/\delta^{2}=1/(2\sigma_{s})^{2}+1/\sigma_{g}^{2}$ and $\Delta(z)=[1+(z/k\sigma_{s}\delta)^{2}]^{1/2}$.
Eqs. (\ref{Fx_cons})-(\ref{Fz_nc}) express the force exerted   on a dipolar particle in the far zone by the field emitted from a GSMS of any state of coherence.

\begin{figure}[h]
\begin{centering}
\includegraphics[width=\linewidth]{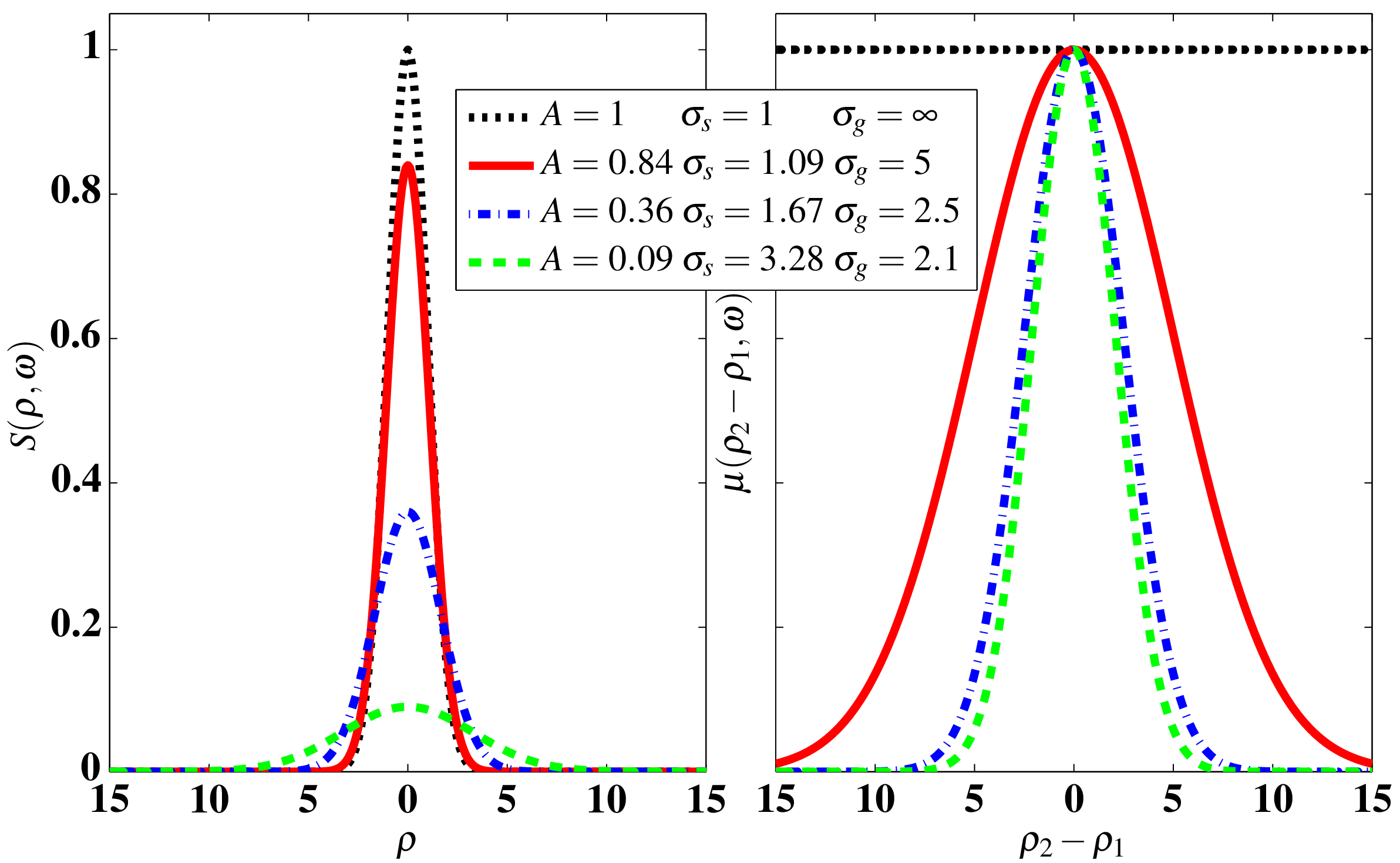}
\par\end{centering}
\caption{(Color online). Spectral density (left) and spectral degree of coherence (rigth) at $z=0$ for different source parameters which generate the same radiant intensity in the far-zone.} 
\end{figure}

\begin{figure}[h]
\begin{centering}
\includegraphics[width=\linewidth]{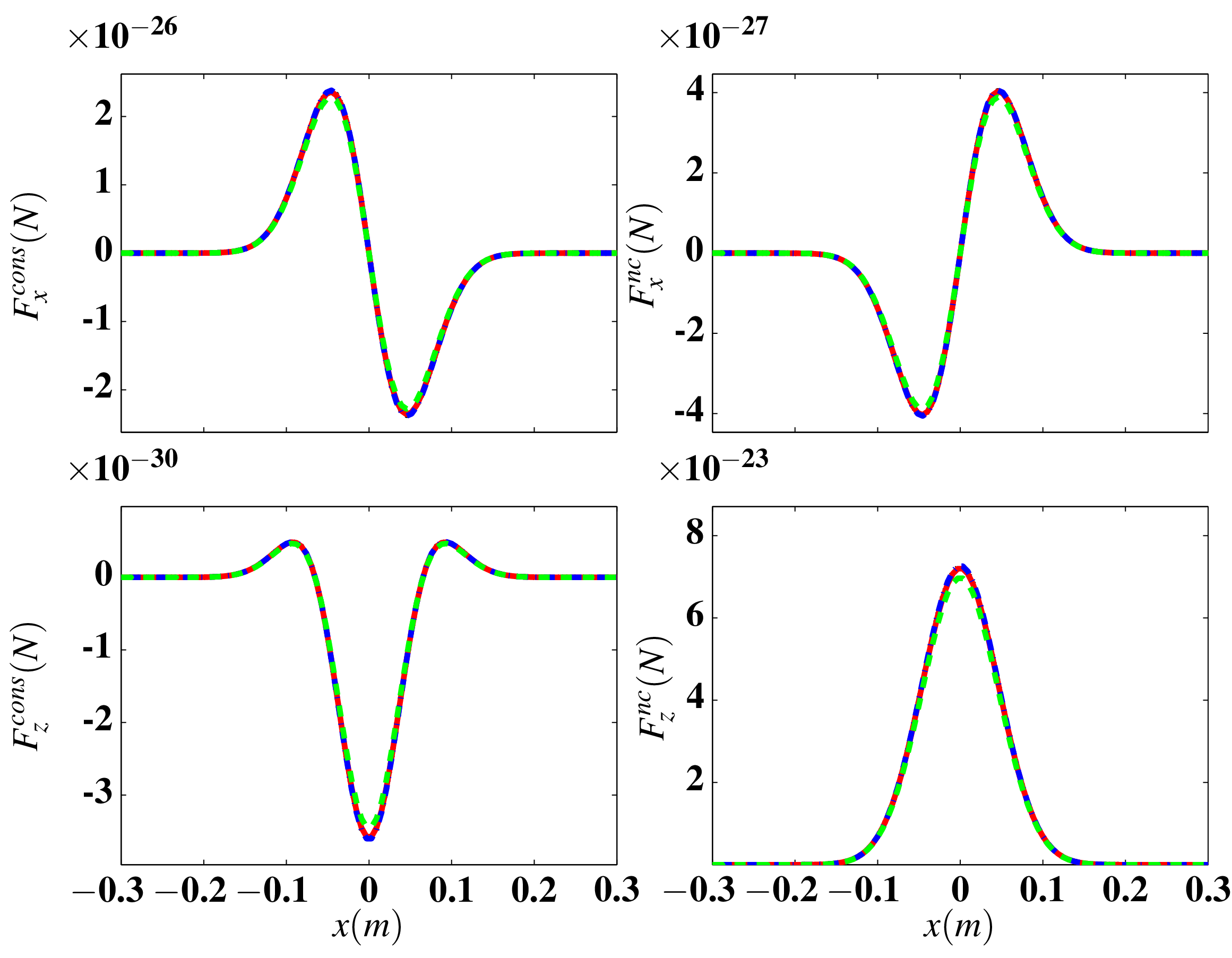}
\par\end{centering}
\caption{(Color online). Conservative (first column) and non-conservative (second column) forces in the  far zone for the same source parameters as in Fig. 1 with $A$ in $mW/ \mu m^2$ and $\sigma_g$ and $\sigma_s$ in $mm$ } 
\end{figure}

Now we address the ET for GSMS \cite{Wolf1978theorem}, and its experimental confirmation  \cite{Gori1979Example}. This theorem establishes that any GSMS will generate the same radiant intensity $J(\theta)=r^2\text{Tr}W_{ij}(\textbf{r},\textbf{r},\omega)$, ($\theta=\rho/z$), as a laser whose spectral density at the plane $z=0$ is $S_l(\boldsymbol{\rho},\omega)=A_l\text{exp}[-\boldsymbol{\rho}^2/(2\sigma_l)]$, if the following conditions are fulfilled 
\begin{align}
\frac{1}{\sigma_{g}^{2}}+\frac{1}{(2\sigma_{s})^{2}}=\frac{1}{(2\sigma_{l})^{2}},\,\,\,\,\,A=\left(\frac{\sigma_{l}}{\sigma_{s}}\right)^{2}A_{l}.
\end{align}
Fig. (1) shows the spectral density and the spectral degree of coherence for the same parameters as in Ref. \cite{Wolf1978theorem}. Any of these source configurations produces exactly the same radiant intensity.
Then, in the context of optical forces from such a partially coherent light, it is natural to ask whether  these fields would trap a particle like a laser beam does.

When the field is beam-like, we can approximate $\Delta(z)\simeq z/(k\sigma_{s}\delta)$. On the other hand, and in order to perform the calculations, we consider a  dipolar latex-like particle with radius $r_0$ and  with relative permittivity $\varepsilon_p$ in the Rayleigh limit ($kr_0\ll1$). The response of the particle is characterized by  the expression for the dynamic electric polarizability which conserves energy on scattering, namely by $\alpha_{e}=\alpha_{0}/\left(1-i\frac{2}{3}k^{3}\alpha_{0}\right)$, 
$\alpha_0=r_{0}^{3}(\varepsilon_{p}-1)/(\varepsilon_{p}+2)$ being the static polarizability. For instance, we choose a wavelength $\lambda=579 \text{nm}$. At this wavelength, for a radius of $r_0=25 \text{nm}$ and for a constant value of $\varepsilon_0=2.25 $, the polarizability has $\text{Re}\alpha_e=4593  \text{nm}^3\gg \text{Im}\alpha_e=17\text{nm}^3$. {\it It should be stressed that much larger particles would lead to similar results providing they may be considered as dipolar, namely, their scattered field be  exactly described by the first electric and magnetic Mie partial waves} \cite{nieto2010optical,JOSA2011}. 

Now we calculate the optical forces in this regime.  Thus now, and in order to answer the latter question posed above, Fig. 2 shows the different contributions (conservative and non-conservative) to the force for the same parameters as in Fig. 1. We see that all  force plots coincide with each other , i.e., \textit{ one does not need a globally spatially coherent source, like a laser beam, as the only light source  capable of building an optical trap. Any partially coherent Gaussian-Schell model source with appropriate spot size and coherence length also does it}. This constitutes the main results of this letter.  For instance, as seen on comparing Figs. 1 and 2, a source with small coherence length $\sigma_g=2.1$ and peak intensity as low as $A = 0.09$ provides the same far zone force as a fully coherent one with a much larger power $A=1$ providing the spot size at the partially coherent source $\sigma_s=3.28$ is larger than the one $\sigma_s=1$ of the fully coherent source.

It should be noticed that given the trade-off  between  $\sigma_g$ and $\sigma_s$ contained in the parameter $4a$, cf. below Eq.(8) and also Eq. (14), there  are infinite  GSMS that yield the same radiant intensity \cite{mandel1995optical}, providing they all lead to the same $4a$ or $\sigma_l$. Thus, the same conclusion may be derived for the optical force induced by their emitted wavefield. On the other hand, it worth remarking that a peak intensity $A$ and width $\sigma_s$ in the spectral density yields an integrated value of this latter magnitude in the source plane: $A\sigma_S^2/2\pi$, which by the Parseval theorem of Fourier transforms is kept in the far-zone. Therefore, a decrease of $A$ while controlling  $\sigma_s$  as mentioned in the latter paragraph, may also lead to lower values of the total total power while maintining the trap essential characteristics. In addition, since the gradient forces push the particle to the peak values of the intensity, this possibility of decreasing $A$ without altering the trap is of crucial importance to minimize the invasive character of the manipulation process.
 
{\it In biophysical experiments, where there is  sample damage produced by high values of the power intensity of the incident beam, this issue acquires vital importance, this equivalence of low peak  intensity  partially coherent fluctuating sources and a high power laser beam, (or in general between different GSMS), as regards the depth and width of the potential well created by the photonic  trap, constitutes a new principle for optical nanomanipulation. Notice that on adequately selecting the source parameters, one can minimize the optical peak intensity and hence the invasiveness, (see the last row of the legend in Fig. 1). As a matter of fact, this peak power can be reduced up to almost two orders of magnitude without varying the effectiveness of the optical trap.} 

The  potential well is usually improved on focusing the emiited light through a thin lens or ABCD system \cite{Friberg1988Imaging}. We have checked this phenomenon when the light is focused by a thin lens, and we get  results with the partially coherent source which are quite similar to those obtained with the coherent beam.

In summary, we have studied the different components of the optical force from a Gaussian Schell model source. it has been demonstrated that in the far-zone such a partially coherent source can produce a force equal to that exerted by a  laser beam; such a force being  larger as  the beam-condition is more  strongly fulfilled. In opposition to this, in the near-field the maximum force corresponds to a minimum force in the far-field. We believe that these results should trigger a renewed interest, as well as experiments, in optical manipulation.
\section{Acknowledgements}
We acknowledge research grants from the Spanish Ministerio de Economia y Competitividad  (MINECO):  CSD2007-00046,  FIS2009-13430-C02-01  and FIS2012-36113-C03-03. J.  M.  A thanks a scholarship from MINECO.

\begin{thebibliography}{10}
\newcommand{\enquote}[1]{``#1''}

\bibitem{Ashkin1970}
A.~Ashkin, \enquote{{Acceleration and trapping of particles by radiation
  pressure},} Phys. Rev. Lett. \textbf{{\bf 24}}, 156--159 (1970).

\bibitem{Ashkin1987science}
A.~Ashkin and J.~Dziedzic, \enquote{Optical trapping and manipulation of
  viruses and bacteria,} Science \textbf{235}, 1517--1520 (1987).

\bibitem{grier2003revolution}
D.~G. Grier, \enquote{A revolution in optical manipulation,} Nature
  \textbf{424}, 810--816 (2003).

\bibitem{ChaumetOL}
P.~C. Chaumet and M.~Nieto-Vesperinas, \enquote{{Time-averaged total force on a
  dipolar sphere in an electromagnetic field},} Opt. Lett. \textbf{{\bf 25}},
  1065--1067 (2000).

\bibitem{Chaumet2002PRL}
P.~C. Chaumet, A.~Rahmani, and M.~Nieto-Vesperinas, \enquote{Optical trapping
  and manipulation of nano-objects with an apertureless probe,} Phys. Rev.
  Lett. \textbf{88}, 123601 (2002).

\bibitem{Kall2002PRL}
H.~Xu and M.~K\"all, \enquote{Surface-plasmon-enhanced optical forces in silver
  nanoaggregates,} Phys. Rev. Lett. \textbf{89}, 246802 (2002).

\bibitem{Arias2002modeling}
J.~R. Arias-Gonz\'alez, M.~Nieto-Vesperinas, and M.~Lester, \enquote{Modeling
  photonic force microscopy with metallic particles under plasmon eigenmode
  excitation,} Phys. Rev. B \textbf{65}, 115402 (2002).

\bibitem{Arias2003attractive}
J.~R. Arias-Gonz\'{a}lez and M.~Nieto-Vesperinas, \enquote{Optical forces on
  small particles: attractive and repulsive nature and plasmon-resonance
  conditions,} J. Opt. Soc. Am. A \textbf{20}, 1201--1209 (2003).

\bibitem{Valdivia2012forces}
F.~J. Valdivia-Valero and M.~Nieto-Vesperinas, \enquote{Optical forces on
  cylinders near subwavelength slits: effects of extraordinary transmission and
  excitation of mie resonances,} Opt. Express \textbf{20}, 13368--13389 (2012).

\bibitem{wang2007effects}
L.~G. Wang, C.~L. Zhao, L.~Q. Wang, X.~H. Lu, and S.~Y. Zhu, \enquote{{Effect
  of spatial coherence on radiation forces acting on a Rayleigh dielectric
  sphere},} Opt. Lett. \textbf{{\bf 32}}, 1393--1395 (2007).

\bibitem{korotkova2009forces}C. Zhao, Y. Cai, and O. Korotkova, 
\enquote{Radiation force of scalar and electromagnetic twisted Gaussian Schell-model beams}, Opt. Express  
{\bf 17},  21472-21487 (2009).


\bibitem{aunon2012photonic}
J.~M. Au{\~n}{\'o}n and M.~Nieto-Vesperinas, \enquote{{Photonic forces in the
  near field of statistically homogeneous fluctuating sources},} Phys. Rev. A
  \textbf{85}, 053828 (2012).

\bibitem{aunon2012opticalforces}
J.~M. Au{\~n}{\'o}n and M.~Nieto-Vesperinas, \enquote{{Optical forces on small
  particles from partially coherent light},} J. Opt. Soc. Am. A \textbf{29},
  1389--1398 (2012).

\bibitem{mandel1995optical}
L.~Mandel and E.~Wolf, \emph{{Optical Coherence and Quantum Optics}} (Cambridge
  U. Press, Cambridge, UK, 1995).

\bibitem{Roychowdhury2005Realizability}
H.~Roychowdhury and O.~Korotkova, \enquote{Realizability conditions for
  electromagnetic gaussian schell-model sources,} Optics Communications
  \textbf{249}, 379 -- 385 (2005).

\bibitem{Wolf1978theorem}
E.~Wolf and E.~Collett, \enquote{Partially coherent sources which produce the
  same far-field intensity distribution as a laser,} Optics Communications
  \textbf{25}, 293 -- 296 (1978).

\bibitem{nieto2004near}
M.~Nieto-Vesperinas, P.~C. Chaumet, and A.~Rahmani, \enquote{{Near-field
  photonic forces},} Phil. Trans. R. Soc. Lond. A \textbf{{\bf 362}}, 719--737
  (2004).

\bibitem{wong2006gradient}
V.~Wong and M.~A. Ratner, \enquote{{Gradient and nongradient contributions to
  plasmon-enhanced optical forces on silver nanoparticles},} Phys. Rev. B
  \textbf{{\bf 73}}, 075416 (2006).

\bibitem{albaladejo2009Scattering}
S.~Albaladejo, M.~I. Marques, M.~Laroche, and J.~J. S{\'a}enz,
  \enquote{{Scattering forces from the curl of the spin angular momentum of a
  light field},} Phys. Rev. Lett. \textbf{{\bf 102}}, 1136021 (2009).

\bibitem{nieto2010optical}
M.~Nieto-Vesperinas, J.~J. S{\'a}enz, R.~G{\'o}mez-Medina, and L.~Chantada,
  \enquote{{Optical forces on small magnetodielectric particles},} Opt. Express
  \textbf{{\bf 18}}, 11428--11443 (2010).

\bibitem{conjug} M. Nieto-Vesperinas and E. Wolf,   \enquote{phase conjugation and symmetries with wave fields in free space containing evanescent components},  J. Opt. Soc. Am. A {\bf 2} 1429-1434 (1985).

\bibitem{nietolibro}
M.~Nieto-Vesperinas, \emph{{Scattering and Diffraction in Physical Optics}}
  (World Science, Singapore, 2006).

\bibitem{wolf2007introduction}
E.~Wolf, \emph{{Introduction to the Theory of Coherence and Polarization of
  Light}} (Cambridge U. Press, New York, 2007).

\bibitem{Gori1979Example}
P.~D. Santis, F.~Gori, G.~Guattari, and C.~Palma, \enquote{An example of a
  Collett-Wolf source}, Optics Communications \textbf{29}, 256 -- 260 (1979).

\bibitem{JOSA2011} M. Nieto-Vesperinas, R. Gomez-Medina, and J. J. Sáenz, \enquote{Angle suppressed
scattering and optical forces on submicrometer dielectric
particles}, J. Opt. Soc. Am. A {\bf 28}, 54–60 (2011).

\bibitem{Friberg1988Imaging}
A.~T. Friberg and J.~Turunen, \enquote{Imaging of Gaussian Schell-model
  sources,} J. Opt. Soc. Am. A \textbf{5}, 713--720 (1988).

\end{thebibliography}

\end{document}